\documentstyle[aps]{revtex}\input{psfig.sty}
\begin{document}
\input epsf
\draft
\title{Sympathetic cooling of an atomic Bose-Fermi gas mixture}
\author{W. Geist, L. You, and T.A.B. Kennedy }
\address{School of Physics, Georgia Institute of Technology \\
Atlanta, Georgia 30332-0430}
\date{\today}
\maketitle
\begin{abstract}
Sympathetic cooling of an atomic Fermi gas by a Bose gas is studied
by solution of the coupled quantum Boltzmann equations for the
confined gas mixture. Results for equilibrium temperatures and
relaxation dynamics are presented, and some simple models
developed. Our study illustrate that a combination of sympathetic
and forced evaporative cooling  enables the Fermi gas to be cooled
to the degenerate regime where quantum statistics, and mean field
effects are important. The influence of mean field effects on the
equilibrium spatial distributions is discussed qualitatively.

\end{abstract}
\section {Introduction \label{sec:intro} }
The magnetic trapping and forced evaporative cooling of alkali
metal vapors has lead to the Bose-Einstein condensation of a number
of different atomic species \cite{bec1}. As a consequence, the
many-body physics of confined, weakly interacting Bose gases are
now amenable to experimental investigation. To date most
experimental and theoretical research on degenerate atomic gases
has focused on bosons \cite{bec2}. Recently, some exotic systems
have been investigated experimentally, these include the Bose
 condensation of spin mixtures of $^{87}$Rb
 using sympathetic cooling \cite{mixtures}, and a report of spin domains
 of $^{23}$Na
 in an optical trap \cite{ketterle}.
There is now a growing interest in the properties of degenerate
atomic Fermi gases
\cite{cornell,stoof,collisions,classic,baranov,marco,rb} and
boson-fermion mixtures \cite{moelmer}, although to date a
degenerate atomic Fermi gas has not been achieved.

In this paper we investigate the process of sympathetic cooling of
an initially non-degenerate Fermi gas to the quantum degenerate
regime using a Bose gas as coolant. Sympathetic cooling of a Fermi
gas to degeneracy is necessary as a result of the suppression of
s-wave scattering between identical fermions in spin symmetric
states. Evaporative cooling of a pure fermion gas trapped in a
single hyperfine state, which depends on rethermalization through
atomic collisions, is thus ineffective at temperatures sufficiently
low that only the lowest few partial waves contribute. The Fermi
gas may instead be cooled by thermal contact with a cold Bose gas,
which may be either already condensed before thermal contact with
the Fermi gas, or evaporatively cooled in its presence. We
investigate the dynamics using the quantum Boltzmann equation (QBE)
for the Bose-Fermi gas mixture. The QBE has been applied to
describe the population dynamics for a Bose gas by a number of
authors who solved it by direct integration \cite{snoke,wally},
trajectory simulation \cite{holland}, and Bird's simulation method
\cite{wu}. The first two sets of authors make the assumption that
the distribution is ergodic at all times \cite{surkov} to simplify
the numerical computation, whereas in Bird's method collision
dynamics are constructed from simulated particle trajectories.
In this paper we solve the QBE in the ergodic approximation
by direct integration of the coupled differential equations which
describe the Fermi and Bose gas distribution functions.

We consider an harmonically trapped gas of $N_f$ fermions with
Fermi temperature given by $E_F\equiv k_B T_F =
\hbar \omega (6 N_f)^{1/3}$, which is in thermal contact
with a similarly confined gas of $N_b$ bosons with condensation
temperature $k_B T_C = \hbar \omega (N_b/1.202)^{1/3}$. By
evaporatively cooling the Bose gas through $T_C$ down to a
temperature $\eta T_C$ ($0< \eta <1$), the Fermi gas will
equilibrate to the Fermi temperature provided $N_f < \eta^3
N_C/7.212$, where $N_C < N_b$ is the number of condensed atoms
remaining. Solution of the QBE gives both the energy and spatial
distributions of the Bose and Fermi gases. An important limitation
of our present treatment is the neglect of mean field effects on
the dynamics. These are potentially important when the Bose and
Fermi gas are degenerate and the mean field energy exceeds the trap
frequency. In a recent paper M{\o}lmer \cite{moelmer} has used a
simple mean field model to study the spatial distributions of a
Bose-Fermi gas mixture at $T=0\;K$. The distributions depend
strongly on the relative sign and magnitude of the boson-boson and
boson-fermion scattering lengths. A complete and numerically
tractable approach to quantum kinetics which incorporates such mean
field effects in addition to the population dynamics, is still
under investigation. A qualitative discussion of the influence of
mean field effects on our results is given in section~\ref{sec3}.

The remainder of this paper is organized as follows. In section
\ref{sec1} we discuss the equilibrium temperatures for an initially
hot
 Fermi gas placed in thermal contact with a cold
Bose gas.
We also derive a simple dynamical model for the thermalization using the assumption
that both gases are described by a Maxwell Boltzmann distribution
at all times. In section~\ref{sec2} we describe a theoretical model
based on the QBE
for the Fermi-Bose mixture and the treatment of forced evaporative cooling.
 In section~\ref{sec3} we present results of solutions of the QBE
which provide information on the kinetic temperature and spatial distributions
of the mixture. Using recent atomic data we also present some results for the
 sympathetic cooling of the $^{40}$K/$^{39}$K
Fermi/Bose potassium isotopes. Finally, in section~\ref{sec4}, we summarize our
conclusions.

\section{ Equilibrium properties and cooling without forced evaporation}\label{sec1}
\subsection{Equilibrium temperature}
If two gases at different initial temperatures are brought into
contact they will rethermalize to a common equilibrium temperature
which can be obtained using conservation of particle number and
energy (over bar indicates average energy)
\begin{eqnarray}
\bar\epsilon_{{\rm tot}}&=&
\bar\epsilon_f(0)+\bar\epsilon_b(0)
\nonumber\\
&=&\bar\epsilon_f(\infty)+\bar\epsilon_b(\infty)
\end{eqnarray}
assuming that there are no losses during relaxation.
For a Bose-Fermi gas mixture the equilibrium temperature $T_{\infty}$
 can be found
numerically using the Bose-Einstein and Fermi-Dirac distribution
functions as follows. Given the number of bosons $N_b$, and
fermions $N_f$, and the initial temperatures $T_b(0)$, $T_f(0)$ we
set $T_f=T_b$ at each iteration step and compute the fugacity $z_f$
for the Fermi gas by solving
\begin{equation}\label{it1}
N_f=\sum_{\epsilon_i}
g(\epsilon_i)(z_f^{-1}e^{\epsilon_i/k_BT_f}+1)^{-1}
\end{equation}
with $g(\epsilon_i)$ the degeneracy of states with energy
$\epsilon_i$. We then compute the mean energy
$\bar\epsilon_f(z_f,T_f)$ for the Fermi gas and determine $T_b$ and
$z_b$ for the Bose gas by the following iteration: set
$\bar\epsilon_b=\bar\epsilon_{{\rm tot}}-\bar\epsilon_f(z_f,T_f)$
and compute $T_b$ from
\begin{equation}\label{it2}
\bar\epsilon_b=\sum_{\epsilon_i}\epsilon_i
g(\epsilon_i)(z_b^{-1}e^{\epsilon_i/k_B T_b}-1)^{-1}.
\end{equation}
Then compute $z_b$ by solving
\begin{equation}\label{it3}
N_b=\sum_{\epsilon_i}
g(\epsilon_i)(z_{b}^{-1}e^{\epsilon_i/k_BT_b}-1)^{-1}
\end{equation}
and repeat this iteration until $[N_b(z_b,T_b)-N_b]/N_b<10^{-7}$
and
$[\bar\epsilon_b(z_b,T_b)-\bar\epsilon_b]/\bar\epsilon_b<10^{-7}$.
We then set $T_f=T_b$ and repeat the procedure until
$(T_f-T_b)/T_f<10^{-4}$.

 For temperatures $T_f \gtrsim 0.5 T_F$ and $T_{\infty} < T_C$ an
approximate equation can be obtained using a classical Maxwell
Boltzmann distribution for the Fermi gas and a Bose distribution
with $z_b=1$ for the Bose gas. This yields
\begin{eqnarray}\label{approxt1}
g_4(1) \frac{T_{\infty}^4}{T_F^4} +\frac{T_{\infty}}{6T_F}=
g_4(1)\frac{T_b^4(0)}{T_F^4} +\frac{T_f(0)}{6T_F}.
\end{eqnarray}
with the Bose-Einstein function $g_4(1)\approx 1.082$. If
$T_{\infty}>T_C$ we can approximate the Bose-Einstein distribution
with a Maxwell-Boltzmann distribution to obtain an explicit
expression for the temperature
\begin{equation}\label{approxt2}
T_{\infty}=\frac{ N_f T_{f}(0)+ N_b T_{b}(0)} { N_f+N_b}.
\end{equation}
In Fig. \ref{fig0} we show the equilibrium temperature as a
function of the initial Fermi gas temperature scaled in units of
$T_F$. The Bose gas consists of $10^6$ atoms at initial temperature
$T_b(0)= 0.1 T_F(0)$, where $T_F$ depends on the number of fermions
which is varied between $10^4$ and $10^6$. In terms of the BEC
temperature the initial temperature of the Bose gas lies in the
range $0.02T_C < T_b(0) < 0.2T_C$. The full numerical results agree
well with approximation Eq.~(\ref{approxt1}) for $N_f=10^3,10^4$
and $10^5$ in the temperature range considered. This is shown by
the curves at the bottom of the figure which almost overlap the
approximate solution. The equilibrium temperature $T_{\infty}$
stays below the critical temperature $T_C$ and in agreement with
Eq.~(\ref{approxt1}) $T_{\infty}/T_F$ does not depend on the number
of fermions. If we increase the number of fermions or alternatively
the initial Fermi gas temperature, the Bose gas will be heated
above $T_C$ and the equilibrium temperature shows the linear
dependence on $T_f(0)$ as  derived in ~(\ref{approxt2}). From our
numerical studies we find that Fermi statistics become significant
only for $T_f \lesssim 0.5 T_F$ (see also \cite{Butts}), which in
general can be reached with additional evaporative cooling of the
gas mixture.

\subsection{Thermalization of a two-component mixture}
We consider Bose and Fermi gases brought into thermal contact in a
confining potential, at different initial temperatures $T_b(0)$ and
$T_f(0)$, respectively. We are primarily interested here in a
regime prior to a stage of forced evaporative cooling, in which the
Bose gas as well as the Fermi gas is non-degenerate. The
thermalization may result in a significant alteration in the
energy and spatial distribution of fermionic atoms in the trap.
A simple dynamical description
of the thermalization can be derived using the classical Boltzmann
equation in the ergodic approximation. The mean energy of component
$k = (b,f)$ is given by
\begin{eqnarray}
\bar\epsilon_{k}(t) = \int \epsilon \; {\mathcal F}_k(\epsilon,t)\;
\rho(\epsilon) d\epsilon,
\end{eqnarray}
where ${\mathcal F}_k(\epsilon,t)$ is the ergodic distribution
function for component $k$, and $\rho(\epsilon) =
\epsilon^{2}/2(\hbar \omega)^3$ is the density of states for an
harmonic trap of mean frequency
$\omega=(\omega_x\omega_y\omega_z)^{1/3}$. For simplicity we
asssume that both components see the same trapping potential.

The time dependence of the energy of the Fermi gas is then obtained
by integrating the Boltzmann equation \cite{snoke} over all
energies
\begin{eqnarray}
\label{energy}
\frac{d\bar E_f}{dt}&=&\frac{1}{\tau_0}
\frac{1}{(\hbar\omega)^5}
\int dE_1\int dE_2
\int dE_3\int dE_4 E_{{\rm min}}^2/2 \; E_1
\delta(E_1+E_2-E_3-E_4)
\nonumber\\
&&
\left[
{\mathcal F}_f(E_4,t){\mathcal F}_b(E_3,t)- {\mathcal F}_f(E_2,t)
{\mathcal F}_b(E_1,t)\right],
\end{eqnarray}
where all energies written with upper case E are dimensionless,
i.e, $E \equiv \epsilon / \hbar \omega$, $E_{{\rm min}}
\equiv {\rm min}\{E_1,E_2,E_3,E_4\}$, and the
natural timescale $\tau_0$ is given in terms of the Bose-Fermi {\it
s}-wave collision cross section $\sigma_{bf}$ by $ 1 / \tau_0
\equiv (\hbar\omega)^2 m\sigma_{bf} / \pi^2\hbar^3$. We have dropped
the term which represents collisions between fermionic atoms
assuming that the {\it p}-wave contribution is neglibible at the
low energies under consideration.

To get some qualitative insight into the thermalization dynamics we
need to further simplify the model. We assume that the component
distribution functions are Boltzmann-like at all times, and
parametrized by time dependent fugacity $z_{k}(t)$ and
dimensionless  temperature  $\bar T_k(t) \equiv k_B T_k(t)/
\hbar\omega$ as
follows,
\begin{eqnarray}
{\mathcal F}[E,z_k(t),\bar T_k(t)] = z_k(t) e^{-E/\bar T_k(t)}.
\end{eqnarray}
The average energy and particle number for a
trapped Maxwell-Boltzmann distribution are given by
\begin{eqnarray}
\bar E_k &=&\frac{z_k}{2}
\bar T_k^4\int dx x^3\;e^{-x}, \\
 N_k &=& \frac{z_k}{2}\bar T_k^3\int dx x^2\;e^{-x}
\end{eqnarray}
from which it follows that $z_k =  N_k/\bar T_k^3$ , and $\bar E_k
= 3  N_k \bar T_k $. The fugacity of component $k$ can therefore
be eliminated in terms of the particle number and mean energy.
Another simplification which results from the approximation is that
the scattering of two bosons does not alter the average energy of
the Bose gas.

It is convenient to write the average energy in terms of a
dimensionless temperature, thus
\begin{eqnarray}\label{tf}
\frac{d\bar T_f}{dt}&=&\frac{N_b r^3}{3 \tau_0} \int dx_1\int dx_2
\int dx_3\int dx_4 \frac{x_{{\rm min}}^2}{2}x_1
\delta(x_1+x_2-x_3-x_4)
\left( e^{-x_4}e^{-r x_3}- e^{-x_2}e^{-r x_1}\right)
\nonumber\\
&=&\frac{ N_b }{3 \tau_0} P(T_f / T_b),
\\
\frac{d\bar T_b}{dt}&=&-\frac{ N_f}{3 \tau_0 }\;\; P(T_f / T_b),
\end{eqnarray}
where $r \equiv \bar T_f /\bar T_b$, and the rational function
\begin{equation}
P(r)=\frac{1+3r+2r^2-2r^3-3r^4-r^5}{(1+r)^5}.
\end{equation}
The equilibrium temperature $\bar T_\infty$ is obtained from $d\bar
T_f/d\bar T_b=- N_b/ N_f$. Integrating from $t=0$ to $\infty$, we recover
 equation ~(\ref{approxt2}). The analysis shows that the
timescale for relaxation is approximately $
\bar T_b \tau_0/ N_f $.
In Fig. \ref{fig1} we compare the predictions of this simple
dynamical model, with the full dynamics of the QBE as discussed in
 the following section. The solutions show
that the time scale that describes  equilibration is on the order
of $\tau_r\equiv \bar T_b\tau_0/ N_f \approx 0.043\tau_0$, as
predicted by Eqs.~(\ref{tf}) and (13).

If forced evaporative cooling is applied to the gas mixture, atoms
from the hot energy tail are removed. The rate of evaporation must
be chosen to be much smaller than the relaxation rate of the gas as
discussed above. Another way to determine the relaxation time is to
calculate the initial mean collision rate $\gamma$ using the
Boltzmann collision integral. Consider the collisions of ``test"
atoms, with energies $E_1$, with atoms of energy, $E_2$, into final
states with energies $E_3$ and $E_4$. One sums over all initial
energies $E_1$, and divides by the number of particles to get the
energy averaged rate per particle
\begin{eqnarray}\label{scatter}
\gamma(\bar T,N)&=& \frac{1}{\tau_0}\frac{N}{\bar T}
\int  dx_1 dx_2 dx_3 dx_4
\delta(x_1+x_2-x_3-x_5)
\frac{x_{{\rm min}}^2}{2} e^{-x_1}e^{-x_2}\\
\nonumber
&=& \frac{N}{2\tau_0\bar T},
\end{eqnarray}
in qualitative agreement with the simple model above. We note, however,
that the relaxation rate we have discussed here should not be regarded as the
 characteristic
time scale for condensation
\cite{smerzi,anglin,li}.

\section{Quantum Boltzmann Equation and forced evaporation}\label{sec2}
\subsection{Quantum Boltzmann equation}
The QBE for a harmonically confined Bose gas has been discussed
elsewhere within the ergodic approximation \cite{holland,gard}. The
QBE for an interacting  two component Bose-Fermi mixture in the
harmonic trap can be written ($\tau = t/\tau_0$)
\begin{eqnarray}\label{bos1}
g(E_i)\frac{db_{E_i}}{d\tau}&=&
\alpha_b \sum_{E_j,E_k,E_l}
\delta_{E_i +E_j, E_k ,E_l}
 g(E_i,E_j,E_k,E_l)
\nonumber\\
&&\left[b_{E_k}b_{E_l}(1+b_{E_j})(1+ b_{E_i})-b_{E_i} b_{E_j}(1+
b_{E_k})(1+b_{E_l})\right]+
\nonumber\\
&& \sum_{E_j,E_k,E_l}
\delta_{E_i +E_j, E_k+E_l}
g(E_i,E_j,E_k,E_l)
\nonumber\\
&&\left[ b_{E_k}f_{E_l}(1-f_{E_j})(1+b_{E_i})-
b_{E_i}f_{E_j}(1-f_{E_l})(1+ b_{E_l})\right],
\end{eqnarray}
\begin{eqnarray}\label{ferm1}
g(E_i) \frac{df_{E_i}}{d\tau}&=&
\alpha_f\sum_{E_j,E_k,E_l}
\delta_{E_i +E_j, E_k
+E_l}\;g(E_i,E_j,E_k,E_l)
\nonumber\\
&&\left[f_{E_k}f_{E_l}(1-f_{E_j})(1- f_{E_i})-f_{E_i} f_{E_j}(1-
f_{E_k})(1-f_{E_l})\right]+
\nonumber\\
&& \sum_{E_j,E_k,E_l}
\delta_{E_i +E_j, E_k+E_l}\;
g(E_i,E_j,E_k,E_l)
\nonumber\\
&&\left[ b_{E_l}f_{E_k}(1-f_{E_i})(1+b_{E_j})-
b_{E_j}f_{E_i}(1-f_{E_k})(1+ b_{E_l})\right],
\end{eqnarray}
where $b_{n}$ and $f_{n}$ are the number of
 Bose and Fermi atoms in state $E_n$.
 The collision matrix elements are approximately given by
\begin{eqnarray}
g(E_i,E_j,E_k,E_l)=g(E_{{\min}}),
\end{eqnarray}
 with $g(E_n)$ the degeneracy of energy level
$E_n$, and $E_{{\rm min}}$ is the minimum energy of all four
energies involved in the scattering process, as defined earlier.
Although this approximation is not quantitatively accurate for the
lowest several states of the trap, it is sufficient to illustrate
the main qualitative features of sympathetic cooling. In an
isotropic trap the degeneracy of the energy state
$\epsilon_n=\hbar\omega (n-1),\;n=1,2\cdots$, is $g(E_n)=n(n+1)/2$.
The coefficients $\alpha_b=\sigma_{bb}/ \sigma_{bf}$ and
$\alpha_f=\sigma_{ff}/\sigma_{bf}$, give the ratios of the cross
sections for boson-boson and fermion-fermion scattering,
respectively, to the boson-fermion cross section. Exchange symmetry
leads to $\alpha_f = 0$ since the {\it s}-wave cross section
vanishes for identical fermions. Of course, this is the reason we
must employ sympathetic cooling with a Bose gas refrigerant.
\subsection{Forced evaporative cooling}
Evaporative cooling in magnetic traps is performed by inducing
transitions to untrapped states with a radio-frequency field. This
is modeled here by the following procedure. Particles that are
scattered into states with energy larger than the time varying
cut-off energy $E_{{\rm cut}}(\tau)$ are lost. The latter is a
given decreasing function of time in the case of forced evaporative
cooling. A particle may be scattered into a state with energy $E_i
< E_{{\rm cut}}(\tau)$,
 by two-body scattering of atoms in states with energies
  $E_k$ and $E_l$, i.e.,
$E_k,\;E_l \rightarrow E_i,\;E_j$, in which $E_j > E_{{\rm
cut}}(\tau)$, so that one particle is lost from the trap. Similarly
a particle in energy level $E_i$ can be scattered out of this
energy level, $E_i,\;E_j
\rightarrow E_k,\;E_l$, resulting in one particle
loss from the trap when either
 $E_k > E_{{\rm cut}}(\tau)$, or $E_l > E_{{\rm cut}}(\tau)$.
Explicitly,\\
 (a) The gain process for energy level $E_i$ \\
 \begin{eqnarray}
 g(E_i)\frac{dn_{E_i}}{d\tau}&=&
\alpha\; \sum_{E_j > E_{{\rm cut}}(\tau)>E_k,E_l}
^{E_{j} < 2 E_{{\rm cut}}(\tau)}
\delta_{E_i +E_j, E_k ,E_l}
 g(E_i,E_j,E_k,E_l)
 n_{E_k}n_{E_l}(1 \pm
n_{E_i}) .
 \end{eqnarray}
(b) The loss process for energy level $E_i$\\
 \begin{eqnarray}
 g(E_i)\frac{dn_{E_i}}{d\tau}&=&
-2\alpha \;\sum_{E_k > E_{{\rm cut}}(\tau)>E_j,E_l}
^{E_{k} < 2 E_{{\rm cut}}(\tau)}
\delta_{E_i +E_j, E_k ,E_l}
 g(E_i,E_j,E_k,E_l)
 n_{E_i} n_{E_j}(1 \pm n_{E_l}),
 \end{eqnarray}
where $n_{E_i}$ denotes the distribution function for Fermi or Bose
atoms with energy $E_i$, as appropriate.

The kinetics of forced evaporative cooling is modeled by adding
these terms  to the QBE, Eqs.~(\ref{bos1}) and (\ref{ferm1}), which
include all two-body collision processes between initial and final
states $i,j,k$, and $l$ with energies $E_i ,E_j, E_k ,E_l < E_{{\rm
cut}}(\tau)$
 conserving the total number of particles in the trap.

\section{Results and discussion}\label{sec3}
In this section we illustrate the dynamics of sympathetic cooling
of Bose-Fermi gas mixtures, through their energy, state and spatial
distribution functions. The spatial distributions of confined
degenerate Bose and Fermi atomic gases are quite different. An
ideal Bose condensate has a size determined by the quantum width of
the trap ground state $l = \sqrt{\hbar/2M \omega}$, whereas the
size of a Fermi gas is governed by the Fermi width $R_F = (E_F/2M
\omega^2)^{1/2}$, which scales as $R_F \sim N_f^{1/6} \; l$, as a
result of the Pauli exclusion principle. For an interacting Bose
gas, with positive scattering length, the condensate is larger than
$l$, and for strongly condensed gases its size can be estimated
using mean field theory in the Thomas-Fermi approximation
\cite{eddi}. Mean field effects, which can be significant well
below the condensation temperature, are not included in our model.
These may be important in the final stages of cooling if the Bose
gas is already strongly condensed at this stage. Our illustrations
of the spatial distributions of both Fermi and Bose gas employ the
universal scaling described by Butts and Rokhsar \cite{Butts}, who
showed that for a harmonically trapped ideal Fermi gas at $T = 0$
\begin{eqnarray}\label{nrf}
n_f(r) = \frac{N_f}{R_F^3} \frac{8}{\pi^2}
\left[ 1 - \left(\frac{r}{R_F}\right)^2 \right]^{3/2}.
\end{eqnarray}
It should be remembered that with evaporative cooling
 the number of fermionic and bosonic
atoms is a time dependent variable, and therefore so is $R_F$. In
the figures we always scale with respect to the instantaneous value
of $R_F$.

In Figs. \ref{fig1} and \ref{fig2} we present the rethermalization
 of a non-degenerate Fermi gas immediately after it is
placed in thermal contact with a Bose gas which is initially at the
Bose condensation temperature $T_C$. The calculation is performed
by numerical integration of the QBE without any forced evaporative
cooling. In Fig. \ref{fig2} the temperature of the Fermi gas alters
considerably over a timescale $
\tau_0\bar{T}_f/N_b$ [Eq. (\ref{tf})].
Initially a hot tail of atoms extends to the trap extremities and
during the early stages of equilibration the Fermi gas distribution
deviates significantly from a Fermi-Dirac distribution which is fit
to the average energy and particle number. The gas then
equilibrates to a non-degenerate state as can be seen from
inspection of the peak of the spatial distribution function,
$n_f(r=0) \ll 1$, \cite{Butts}.
 The Bose gas has one
hundred times more particles than the Fermi gas, and completely
envelopes the Fermi gas at all times. Fig. \ref{fig2} compares the
simple model of thermalization discussed in the last section with
the QBE. The model is very good in the early stages, but the
agreement deteriorates in the intermediate regime before steady
state is achieved.

In Figs. \ref{fig3} and \ref{fig4} we consider the forced
evaporative cooling of both gases. In contrast to Figs. \ref{fig1}
and \ref{fig2} there are $10^5$ bosons and $10^4$ fermions,
initially. The forced cooling begins after the initial
equilibration stage during which the boson temperature increases
(Fig. \ref{fig4}), following thermal contact of the two gases.
 The evaporative cooling time scale is chosen to be $\tau_0$,
which is much longer than both the relaxation time scale of the one
component Bose gas [Eq.~(\ref{scatter})] and the relaxation time
scale of the Fermi gas with the Bose gas [Eq.~(\ref{tf})]. The Bose
gas energy distribution shows the formation of the condensate and
the corresponding spatial distribution contracts to that of the
condensate with a small thermal component. The degenerate Fermi gas
is then exposed and its spatial distribution is close to the zero
temperature limit [Eq.~(\ref{nrf})] which has a maximum density at
the trap center $n_f(0)R_F^3/N_f=8/\pi^2\approx 0.81$. The inset
shows the state occupancy for the Fermi distribution with the
characteristic smearing of the Fermi surface at finite temperature,
and near unit occupancy for  low lying levels.  It is interesting
to note that evaporative cooling of the Fermi gas
 still proceeds at later times when the
spatial overlap between fermions and bosons is mainly in a small
region at the center of the trap where the condensate is located.
The collisions which result in cooling involve orbits of hot
fermions through the trap center where they collide with cold
condensed bosons. At this stage evaporation mainly results in
depletion of fermions as can be seen in Fig. \ref{fig4}. We also
simulate the case when the evaporative cooling involves only the
loss of Bose atoms from the trap. In Figs. \ref{fig5} and
\ref{fig6} we consider the same initial conditions as for Figs.
\ref{fig3} and \ref{fig4}, but only ramp down the cut-off energy
for the bosons. The results are qualitatively the same as in the
former case where both Bose and Fermi gas particles evaporate,
except that we end up with more particles left in the Fermi gas. As
mentioned earlier we have not included the effect of the bosonic
mean field which can alter the spatial distributions of each
component depending on the mean field strength, the ratio of the
scattering lengths, and the particle numbers of both components
\cite{moelmer}.

A possible experimental scenario for sympathetic cooling of a
Bose-Fermi mixture involves two isotopes of potassium. Recent
calculations predict the {\it s}-wave scattering length for the
bosonic isotope $^{39}$K to be $a(^{39}{\rm K}) \equiv a_b =4.3$
(nm) with corresponding cross section $\sigma_{bb}=8\pi^2a_b^2$ and
the {\it s}-wave scattering length between $^{39}$K and the
fermionic isotope $^{40}$K to be $a(^{40}{\rm K}-^{39}{\rm
K})\equiv a_{bf} = 2.5$ (nm) with cross section
$\sigma_{bf}=4\pi^2a_{bf}^2$ \cite{potassium}. Using $M(^{40}{\rm
K})=6.6\times 10^{-26}kg$  and $\omega=400\; $(Hz) sets the time
scale $\tau_0 = 1254 \;({\rm s})
\approx 20$(min). Mean field effects become important when the mean
field strength $E_{{\rm mean}}=4\pi\hbar^2a_b/ (M(^{40}{\rm K})\bar
n_0)$ is of the order of the level spacing $\hbar\omega$
~\cite{singh}. At the onset of BEC the density profile of an ideal
Bose gas is almost Gaussian and the mean density of the ground
state becomes $\bar n_0
= N_0/\pi^{3/2}l^3$. The ratio of the mean field strength to
the trap energy is then
\begin{equation}
\gamma=\frac{E_{{\rm mean}}}{\hbar\omega}=
\sqrt{\frac{1}{\pi}}
\frac{a_b}{l}N_0=1.72\times10^{-3}N_0 .
\end{equation}
Further discussions of the influence of mean-field effects on our
results is given below. In Figs. \ref{fig7} and \ref{fig8} we
present an example of evaporative cooling for a spin polarized
$^{40}{\rm K}-^{39}$K mixture of $10^6$ bosons and $10^5$ fermions
at initial temperatures $T_b=T_C\equiv 0.3$ ($\mu$K) and
$T_f=7.2T_F\equiv1.8$ ($\mu$K) and cut-off energy $E_{{\rm
cut}}(0)= 1000$. We estimate the boson scattering rate using Eq.~
(\ref{scatter}) which yields $\approx 5\times 10^3/\tau_0$. From
$\tau=0$ until $\tau=0.04$ the cut-off energy remains at $E_{{\rm
cut}}=1000$. During the thermalization the Bose distribution
completely envelopes the Fermi distribution which deviates
significantly from its equilibrium distribution. As one can see
from Fig. \ref{fig8} the number of fermions almost remains the same
whereas a large number of bosons are evaporated. After $\tau=0.04$
the cut-off energy is ramped down exponentially with rate
$\gamma_{{\rm evap}}=100/\tau_0$ until $\tau=0.064 \cong 80.25$
(s). At the early stages of the forced evaporation the number of
bosons decreases until most of the bosons are in the condensate and
evaporation mainly leads to depletion of fermions. The simulation
shows that the Fermi gas can be cooled to a temperature $T_f\approx
0.1 T_F$ with more than $2\times 10^4$ fermions left in the trap.

In practice there are some additional issues that must be
considered. The isotopes have different mass and magnetic moment,
this means that in general the clouds will be displaced with
respect to one another due to the combined effects of gravity and
the magnetic trapping force \cite{mixtures}. Sympathetic cooling
can only proceed efficiently if good overlap between the gases is
maintained \cite{ketterle2}. Even if we assume this has been
achieved, the difference in magnetic moments will cause the trap
frequencies to be different for the Bose and Fermi gas. For example
if the bosonic $^{39}$K ( $I=7/2$, where I is the nuclear spin) is
polarized in the state $|F=2, M_F=2 \rangle$, and the fermion
isotope $^{40}$K ($I=4$) is polarized in the state $|F=7/2,
M_F=-7/2\rangle$, the trap frequencies would be in the ratio of
$7:9$. In our calculation we assume the trap frequencies and masses
to be identical.

For degenerate Bose and Fermi gases mean field effects can strongly
influence the spatial distributions. Here we discuss how these
effects qualitatively change the stationary distributions presented
above, using the zero temperature model discussed by M{\o}lmer
\cite{moelmer}.
 If the number of bosons is much larger
than the number of fermions then as $a_{bf}$ increases relative to
$a_b$ ($a_{bf},\;a_b > 0$), the fermion distribution is displaced
further and further outside of the central core of the trap
occupied by the bosons. When the particle numbers are similar the
bosons are displaced outside the fermion region in the same limit.
On the other hand, for large negative fermion-boson scattering
length the stationary distributions may be unstable and dynamically
change as a result. In the latter case mean field effects
qualitatively change the nature of the problem, and thus our
results will not apply.

In our simulations the number of bosons is always much larger than
the number of fermions. In this case we can to first order neglect
the influence of the fermions on the the boson spatial distribution.
The boson density is then given by the Thomas-Fermi approximation
\begin{equation}\label{tommy}
n_b(\vec{r})=[\mu -V_{{\rm ext}}(\vec{r})]/a_{b}
\end{equation}
where $V_{{\rm ext}}(\vec{r})= M\omega^2r^2 / 2$ denotes the
harmonic trapping potential, and the chemical potential $\mu$ is
fixed through the condition $\int d\vec r n_b(\vec r)=N_b$.
Explicitly, this yields
\begin{equation}
\mu=[(\frac{m\omega^2}{2})^{3/2}\frac{15}{8\pi}N_b a_b]^{2/5}.
\end{equation}
The bosonic mean-field produces the well known broadening of the
boson density distribution relative to the ground state length of
the trap. The corresponding spatial distribution of fermions can be
found from the equation \cite{moelmer}
\begin{equation}\label{molly}
\frac{\hbar^2}{2M}[6\pi^2n_f(\vec{r})]^{2/3} +
(1-\frac{a_{bf}}{a_{b}}) V_{{\rm
ext}}(\vec{r})+\frac{a_{bf}}{a_{b}}\mu=E_F
\end{equation}
where $E_F$ is determined by $\int d\vec r n_f(\vec r)=N_f$.
Explicitly the density is given by
\begin{eqnarray}\label{nrf2}
n_f(r) = \frac{N_f}{R_F^3} \frac{8}{\pi^2}
\left[ 1 - \frac{a_{bf}}{a_b}\frac{\mu}{E_F}-(1-\frac{a_{bf}}{a_b})
\left(\frac{r}{R_F}\right)^2 \right]^{3/2}
\end{eqnarray}
which may be compared with equation (\ref{nrf}) for an ideal gas.
Clearly the mean field effects cause a broadening in the spatial
distribution of the fermion cloud for $a_{bf} < a_b$. If
$a_{bf}>a_{b}$ the fermions experience an inverted harmonic
oscillator potential near the origin which repells them from this
region. In our results presented above, and in particular for the
example of the potassium isotopes, we always have $a_{bf}<a_{b}/2$.
As a result mean field effects will result in a broadening of both
Bose and Fermi gas spatial distributions, but not a relative
displacement of the clouds.

\section{Conclusion}\label{sec4}
We have discussed the cooling of a confined non-degenerate Fermi
gas to quantum degeneracy using an ultracold Bose gas coolant and
evaporative cooling. Results for the stationary distributions and
dynamics based on solutions of coupled QBE equations for the
Bose-Fermi mixture were presented. These include investigations of
the use of forced evaporative cooling to enhance the degeneracy of
the Fermi gas.
While the QBE does not include mean field effects, which are potentially
important in the quantum degenerate regime, we have discussed
their qualitative effects on the results presented, when
$a_b > a_{bf} > 0$. In this instance mean fields lead to a
broadening of both the Bose and Fermi spatial distributions, but not
a relative displacement of the clouds.

\acknowledgements
We acknowledge support from the NSF.

\newpage

\begin{figure}
\caption{
Equilibrium temperature as a function of the initial Fermi gas
temperature in units of $T_F$. The number of fermions is $N_f=10^3$
(--), $N_f=10^4$ (:), $N_f=10^5$ (.), $N_f=2\times 10^5$ (o),
$N_f=4\times 10^5$ ($\diamondsuit$) , $N_f=10^6$ (+). The solid
lines show the results using Eq.~(\ref{approxt1}) and
Eq.~(\ref{approxt2}) respectively. The number of bosons is
$N_{b}=10^6 $ and the initial temperature is always chosen as $\bar
T_{b}(0)=0.1\bar T_F$ corresponding to $\bar T_{b}(0)=0.019 \bar
T_C$, $0.042 \bar T_C$, $0.09 \bar T_C$, $0.13\bar T_C$, $0.142
\bar T_C $, and $0.19 \bar T_C $. }
\label{fig0}
\end{figure}

\begin{figure}
\caption{The temperature of the Bose gas (solid line)
and of the Fermi gas (broken line) as a function of time. The
dotted line denotes the temperature obtained by solving the
differential equations assuming a Maxwell Boltzmann distributions
with time dependend fugacity and temperature.}
\label{fig1}
\end{figure}

\begin{figure}
\caption{
The Fermi (solid line) and Bose (dotted line) distribution
functions at times $\tau=0$, $0.003$ and $0.016$. The graphs on the
left show the number of atoms as a function of energy, the graphs
on the right the spatial distribution. Initial conditions are
$N_{b}(0)=10^5$, $N_{f}(0)=10^3$, and $\bar T_{b}(0)=\bar T_C=
43.7$, $\bar T_{f}(0)= 5\bar T_F= 81.8$. Both gases are in contact
at $\tau=0$ and then relax to the equilibrium temperature $\bar
T_{\infty}=44$. The broken lines denote the fit to the Fermi
distribution and coincide with full lines at $tau=0$ and
$\tau=0.016$.}
\label{fig2}
\end{figure}

\begin{figure}
\caption{
The Fermi (solid line) and Bose (dotted line) distribution
functions at times  $\tau=0$, 0.9, and 2.0. The graphs on the left
show the number of atoms as a function of energy, the  graphs on
the right  the spatial distribution, the inset the number of
fermions per energy divided by the degeneracy. Initial conditions
are $N_{b}(0)=10^5$, $N_{f}(0)=10^4$, $\bar T_{b}(0)=\bar
T_C=43.7$, $\bar T_{f}(0)=5\bar T_F=186$. From $\tau=0$ until $\tau
=0.04$ the cut-off for the Bose gas is $E_{{\rm cut}}=500$ and for
the Fermi gas $E_{{\rm cut}}=1000$. After $\tau =0.04$ the cut-off
energy is ramped down with a rate $\gamma=1.0$ for both gases
starting at $E_{0}=500$. [$E_{{\rm cut}}(\tau)=e^{-\gamma
\tau}E_{0}$]. The fit to the Fermi distribution is also drawn as a broken line.
The temperatures from top to bottom are $\bar T_{f}(\tau)=5\bar
T_F$, $0.6\bar T_F$ , $0.14\bar T_F$, and $\bar T_{b}(\tau)=\bar
T_C$, $0.5\bar T_C$, $0.1\bar T_C$}\label{fig3}
\end{figure}
\begin{figure}
\caption{
Upper graph: The number of bosons and fermions remaining in the
trap. Lower graph: The temperature of the gases as a function of
time. The dotted lines show the critical temperature for the Bose
gas and the Fermi temperature.}\label{fig4}
\end{figure}

\begin{figure}
\caption{
The Fermi (solid line) and Bose (dotted line) distribution
functions at times  $\tau=0$, 0.9 and 2.2. The graphs on the left
show the number of atoms as a function of energy, the graphs on the
right the spatial distribution, the inset the number of fermions
per energy divided by the degeneracy. Initial conditions are
$N_{b}(0)=10^5$, $N_{f}(0)=10^4$, $\bar T_{b}(0)=\bar T_C=43.7$,
$\bar T_{f}(0)=5\bar T_F=186$. From $\tau=0$ until $\tau=0.04$ the
cut-off for the Bose gas is $E_{bcut}=500$ and for the Fermi gas
$E_{fcut}=1000$. After $\tau
=0.04$ the cut-off energy is ramped down with a rate $\gamma=1.0$
only for the Bose gas starting at $E_{0}=500$. [$E_{{\rm
cut}}(\tau)=e^{-\gamma
\tau}E_{0}$].
The fit to the Fermi distribution is also drawn as a broken line.
The temperatures from top to bottom are $\bar T_{f}(\tau)=5\bar
T_F$, $0.5\bar T_F$, $0.14\bar T_F$, and $\bar T_{b}(\tau)=\bar
T_C$, $0.5\bar T_C$, $0.14\bar T_C$.}\label{fig5}
\end{figure}

\begin{figure}
\caption{
Upper graph: The  number of bosons and fermions remaining in the
trap. Lower graph: The temperature of the gases as a function of
time. The dashed lines show the critical temperature for the Bose
gas and the Fermi temperature.}\label{fig6}
\end{figure}

\begin{figure}
\caption{ Sympathetic cooling of the fermionic Potassium isotope $^{40}$K by the
bosonic isotope $^{39}$K. The Fermi (solid line) and Bose (dotted
line) distribution functions at times $\tau=0$, $3.6\times
10^{-4}$, $4.7\times10^{-2}$, and $6.1\times10^{-2}$. The graphs on
the left show the number of atoms as a function of energy, the
graphs on the right the spatial distribution, the inset the number
of fermions per energy divided by the degeneracy. Initial
conditions are $N_{b}(0)=10^6$, $N_{f}(0)=10^5$, $\bar
T_{b}(0)=\bar T_C=94.1$, $\bar T_{f}(0)=7.2\bar T_F=590.4$. From
$\tau=0$ until $\tau=0.04$ the cut-off for both gases $E_{{\rm
cut}}=1000$. After $\tau =0.04$ the cut-off energy is ramped down
with a rate $\gamma=100.0$. The broken line denotes the fit to the
Fermi distribution . The Fermi gas temperatures from top to bottom
are $\bar T_{f}(\tau)=7.2\bar T_F$, $1.62\bar T_F$, $0.67\bar T_F$,
and $0.14\bar T_F$. The Bose temperature is $\bar T_{b}(\tau)=\bar
T_C$, $1.02\bar T_C$, $0.63\bar T_C$, and $0.12\bar
T_C$.}\label{fig7}
\end{figure}

\begin{figure}
\caption{
Upper graph: The  number of bosons and fermions remaining in the
trap. Lower graph: The temperature of the gases as a function of
time. The dashed lines show the critical temperature for the Bose
gas and the Fermi temperature.}\label{fig8}
\end{figure}


\begin{references}
\bibitem{rb} Millions of $^{82}$Rb atoms have been trapped and cooled
by  Dr. Vieira's group at the Los Alamos National Laboratory.
Similar technique can also be applied to  $^{84}$Rb and $^{86}$Rb
atoms. X. Zhao, (private comuunications, 5/98).
\end{references}
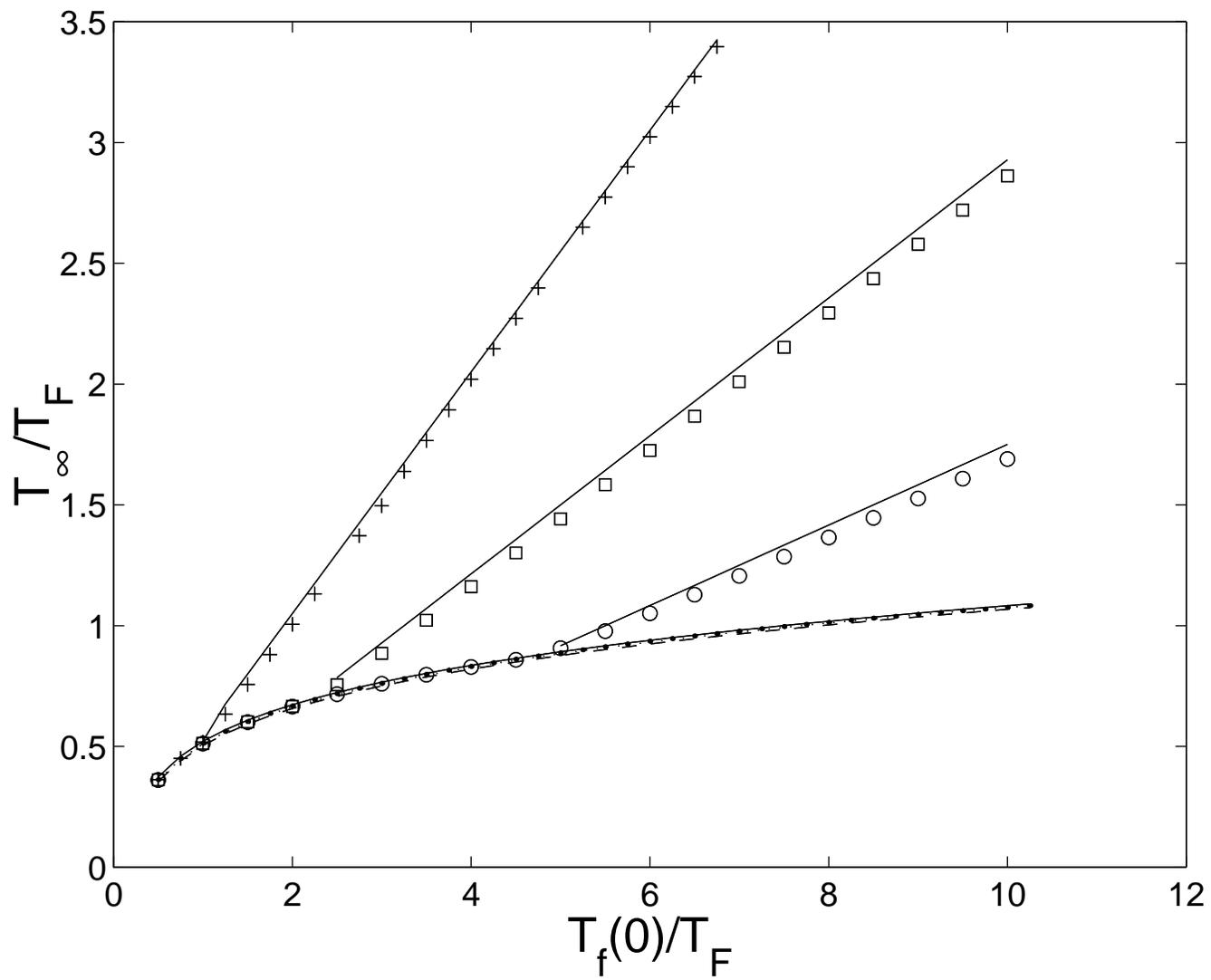
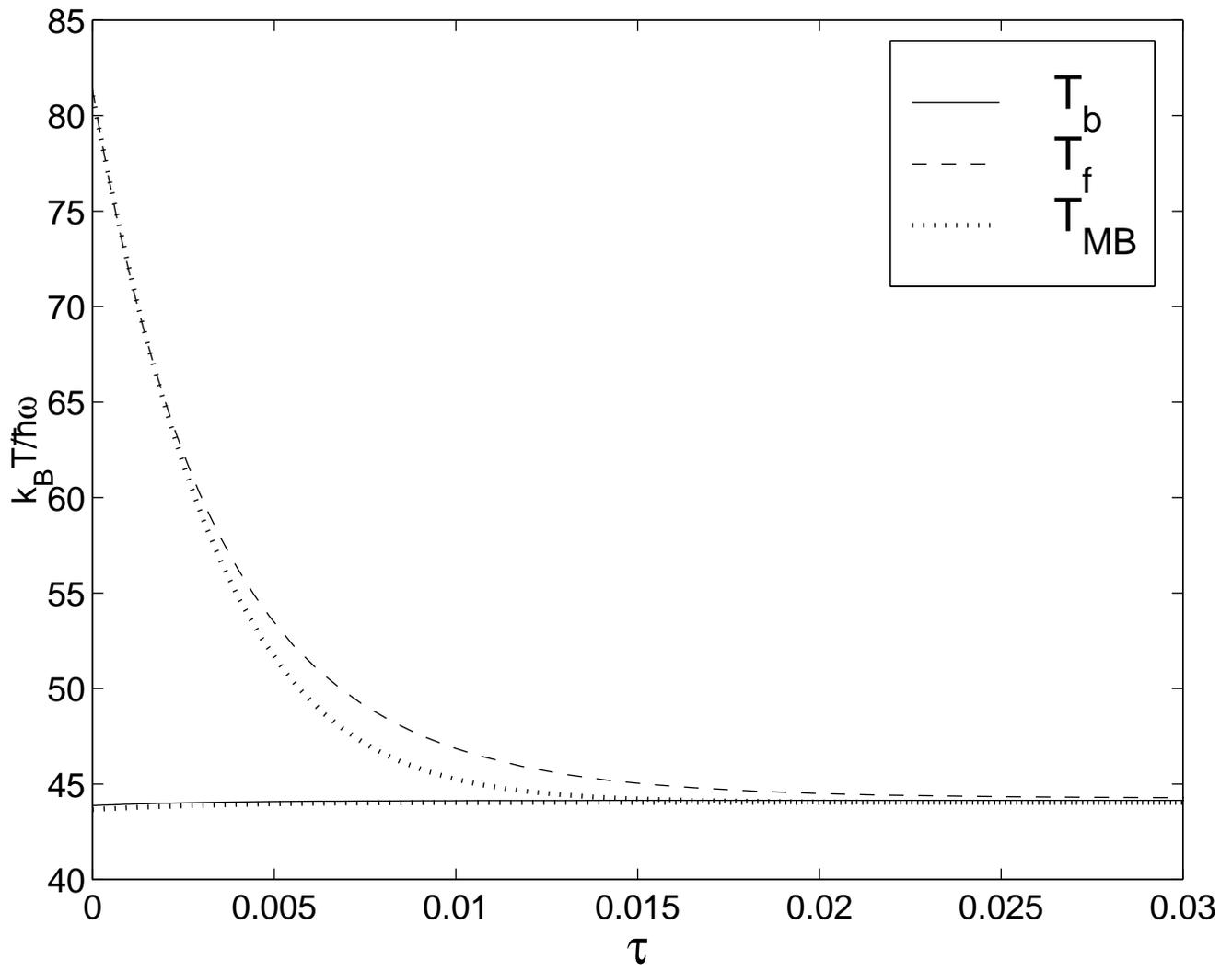
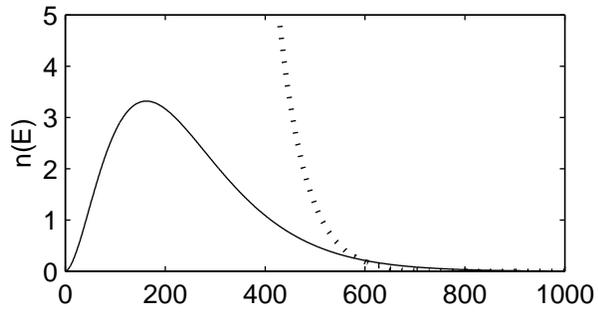
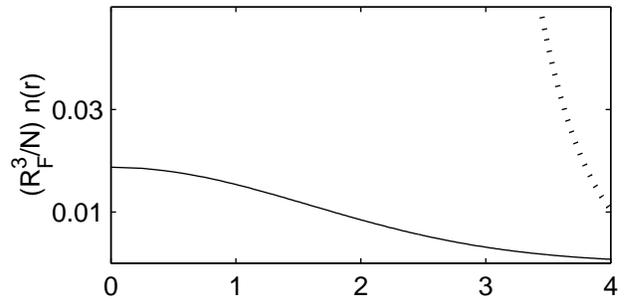
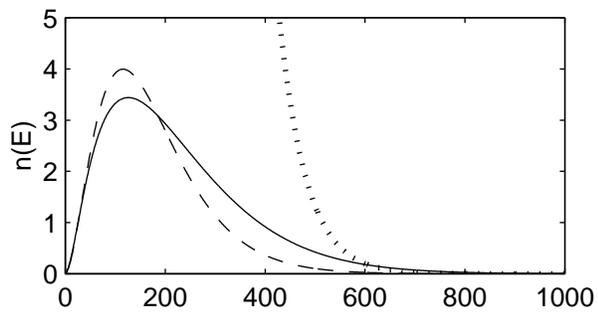
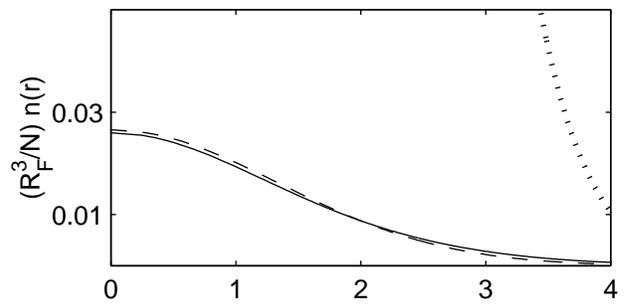
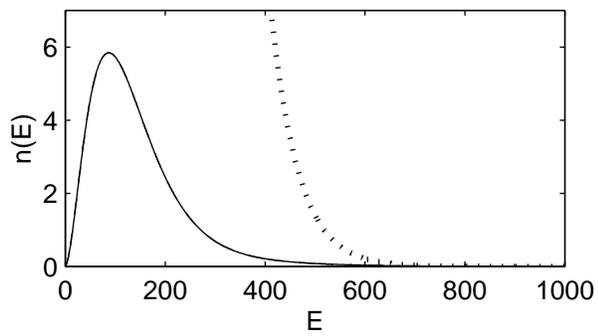
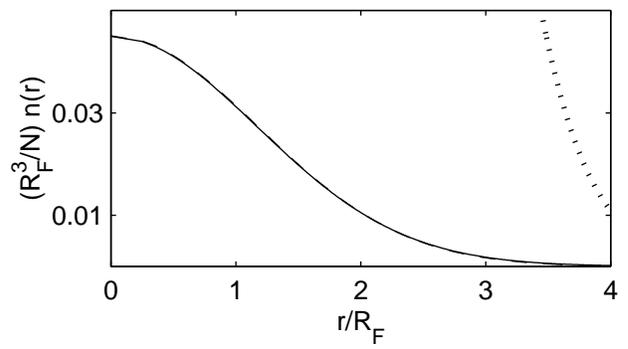
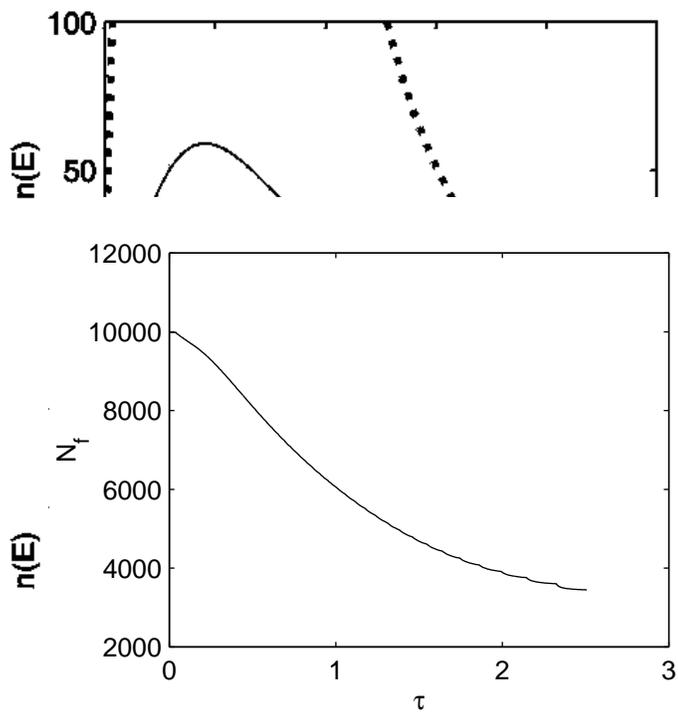
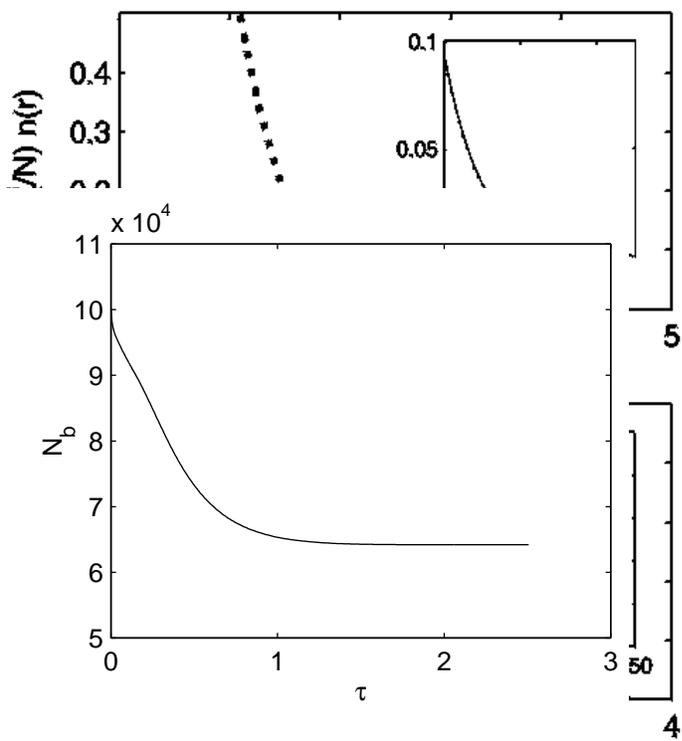
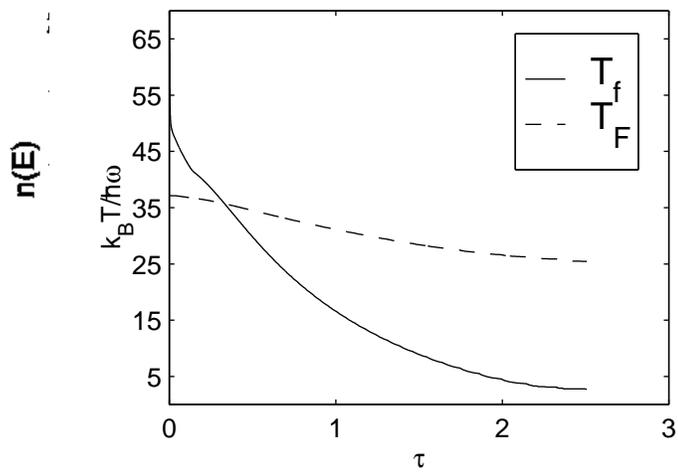
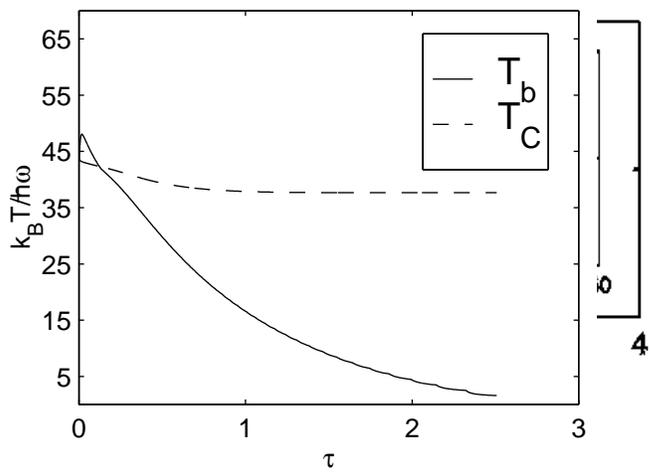
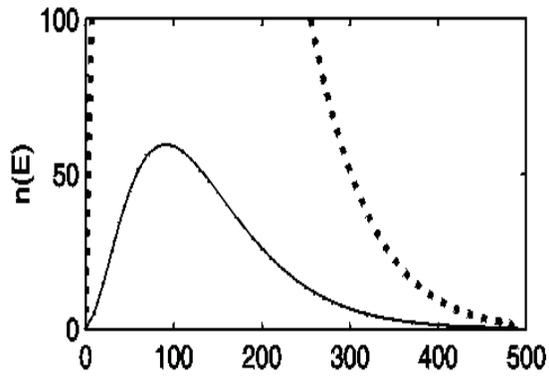
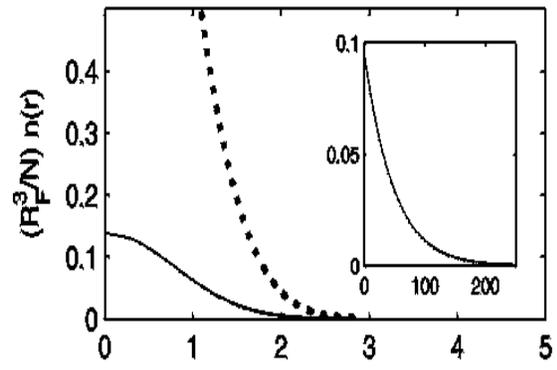
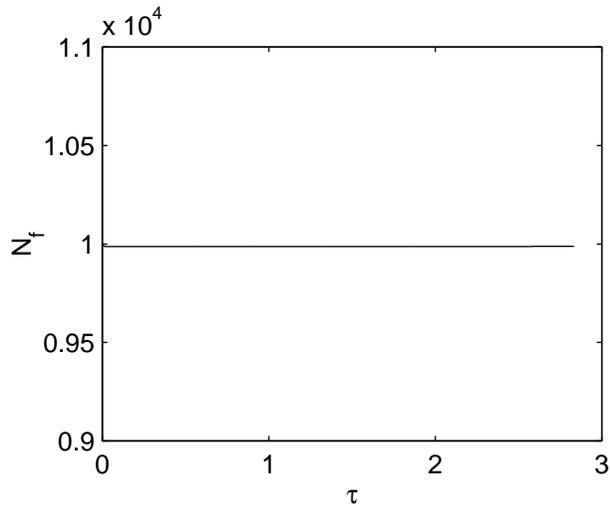
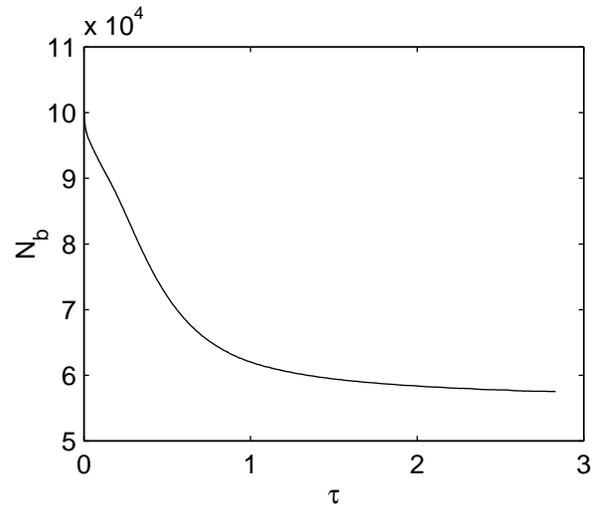
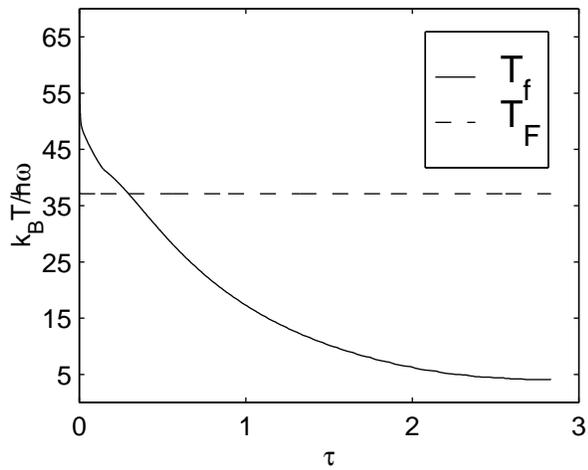
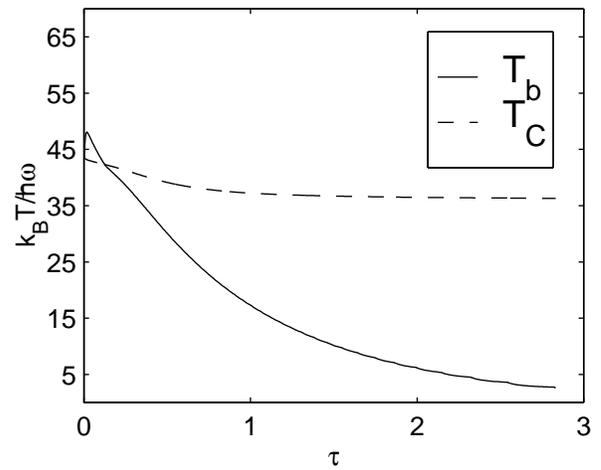
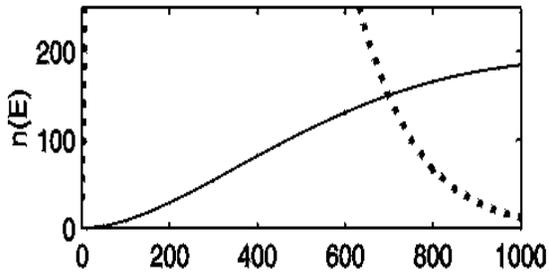
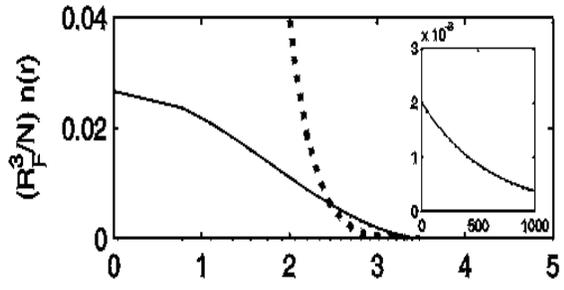
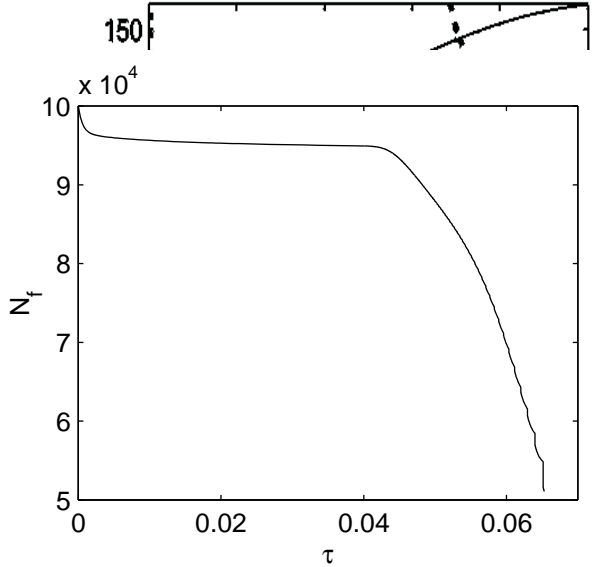
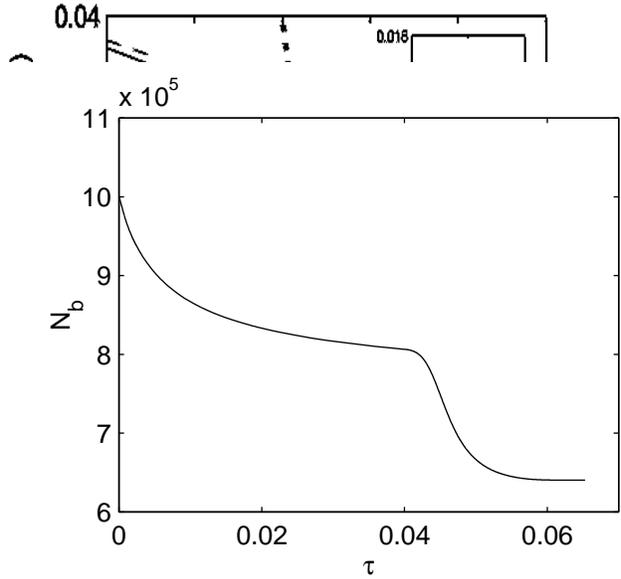
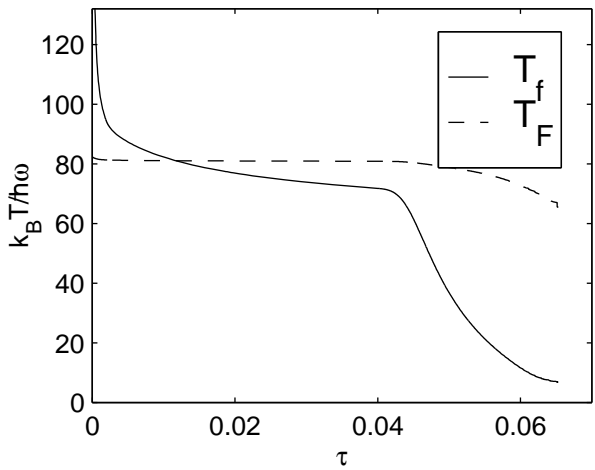
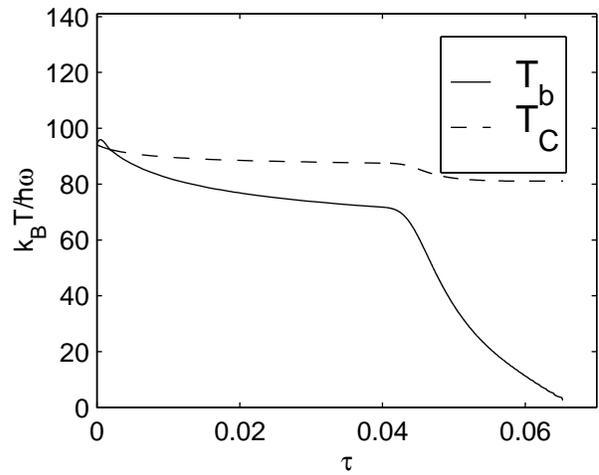
\end{document}